\documentstyle[12pt]{article}
\begin{document}
\hspace{100mm}  UICHEP-TH/99-8
\begin{center}
{\bf    Quasiclassical Analysis of the Three-dimensional
          Schr\"odinger's Equation \\ and Its Solution }  \\
\vspace{3mm}
\rm                     {   M. N. Sergeenko   }       \\
\vspace{2mm}
\it{        The National Academy of Sciences of Belarus,
           Institute of Physics \\ Minsk 220072, Belarus \ and \\ }
\it{  Department of Physics, University of Illinois at Chicago,
                      Illinois 60607, USA }
\end{center}

\begin{abstract}
The three-dimensional Schr\"odinger's equation is analyzed with the 
help of the correspondence principle between classical and 
quantum-mechanical quantities. Separation is performed after 
reduction of the original equation to the form of the classical 
Hamilton-Jacobi equation. Each one-dimensional equation obtained 
after separation is solved by the conventional WKB method.  
Quasiclassical solution of the angular equation results in the 
integral of motion $\vec M^2=(l+\frac 12)^2 \hbar^2$ and the 
existence of nontrivial solution for the angular quantum number 
$l=0$. Generalization of the WKB method for multi-turning-point 
problems is given. Exact eigenvalues for solvable and some 
``insoluble'' spherically symmetric potentials are obtained.  
Quasiclassical eigenfunctions are written in terms of elementary 
functions in the form of a standing wave.  \end{abstract}

\noindent {\bf 1. Introduction }\\

Basic equation of quantum mechanics, the Schr\"odinger's wave 
equation, is usually solved in terms of special functions or 
numerically; for several potentials, the equation is solved exactly 
\cite{He}. The general approach to solve the Schr\"odinger's equation 
for the solvable potentials\footnote {By "solvable" potentials we 
mean those models for which the eigenvalue problem can be reduced to 
a hypergeometric function by a suitable transformation.} is to reduce 
this equation to the equation for hypergeometric function or some 
special function. To do that one needs to find first a special 
transformation for the wave function and its arguments to reduce the 
original equation to the hypergeometric form. After that using 
certain requirements (defined by boundary conditions) to the 
hypergeometric function one can write the corresponding solution for 
the problem under consideration, i.e. the eigenfunctions and the 
corresponding eigenvalues.

This is rather a mathematical approach to solve the eigenvalue 
problem in quantum mechanics; the corresponding methods and solutions 
of the wave equation for some potentials have been developed long 
before the creation of quantum mechanics. There are several features 
of this exact mathematical method that should be clarified from the 
physical point of view. One of them is related to the $S$-wave state.  
The radial Schr\"odinger's equation has no the centrifugal term for 
the orbital quantum number $l=0$. From the physical point of view, 
this means that the problem does not have the left turning point. In 
order for the physical system to have a stable bound state (discrete 
spectrum) two turning points are required (see Ref. \cite{He}, for 
instance). However, solving the radial wave equation for $l>0$, one 
obtains energy eigenvalues for all $l$ including $l=0$. Another 
feature is related to the angular dependence. The angular 
eigenfunction for the ground state, $Y_{00}(\theta,\varphi)=const$, 
i.e. no nontrivial solution exists. Meanwhile, as in the case of 
radial dependence, it might be a function with no zeroes of the type 
of a standing half-wave.

There is another approach to the eigenvalue problem in quantum 
mechanics. This is the quasiclassical method which is well known and 
widely used mainly as the Wentzel-Kramers-Brillouin (WKB) 
approximation \cite{Fro}-\cite{Tr} applicable in the case when the de 
Broglie wavelength, $\lambda =h/p$ ($h=2\pi\hbar$), is changing 
slowly.  In several cases of interest, the WKB method yields the 
exact energy levels, however, its correct application results in the 
exact energy eigenvalues for {\em all} known solvable potentials.

The quasiclassical method is based on the correspondence principle 
between classical functions and operators of quantum mechanics.  The 
correspondence principle is used to derive the wave equation in 
quantum mechanics. In Ref. \cite{SeS} this principle has been used to 
derive the semiclassical wave equation appropriate in the 
quasiclassical region. It was shown that the standard WKB method (to 
leading order in $\hbar$) is the appropriate method to solve this 
equation.

In this work we solve the multi-dimensional Schr\"odinger's equation 
by the quasiclassical method. Unlike known approaches, instead of 
modification of one-dimensional equations obtained after separation, 
we analyze an original multi-dimensional Schr\"odinger's equation and 
reduce it to the equation in canonical form (without first 
derivatives). Separation is performed after reduction of the equation 
obtained to the form of the classical Hamilton-Jacobi equation. We 
show that the main question of the exactness of the quasiclassical 
method is tightly connected with the correspondence principle, i.e.  
the form of the generalized moments obtained after separation of the 
wave equation; the moments obtained after separation have to coincide 
with the corresponding classic generalized moments. The quantization 
condition is written with help of the argument's principle in the 
complex plane that allows us to generalize the quasiclassical method 
for multi-turning-point problems and obtain the exact energy 
eigenvalues for all known solvable potentials and, also, for some 
``insoluble'' problems with more then two turning points.

The quasiclassical method reproduces not only the exact energy 
spectrum for known potentials but has new important features. One of 
the consequences of the quasiclassical solution of the 
multi-dimensional Schr\"odinger's equation is the existence of a 
nontrivial angular solution at $l=0$, $\tilde Y_{00}^{WKB}(\theta, 
\varphi)$, which describes the quantum fluctuations of the angular 
moment. This method allows us to show apparently the contribution of 
quantum fluctuations of the angular momentum into the energy of the 
ground state. \\

\noindent {\bf 2. Exactness of the WKB method }\\

It is well known that the exact eigenvalues can be defined with the 
help of the asymptotic solution, i.e. the exact solution and its 
asymptote correspond to the same exact eigenvalue of the problem 
under consideration. The asymptotic solutions in quantum mechanics 
can be obtained by the WKB method. Therefore the quasiclassical 
method should reproduce the exact energy spectrum.

Intriguing results have been obtained with the help of the 
supersymmetric WKB method (SWKB) \cite{Sukh}-\cite{Kha}, which is a 
modification of the standard WKB quantization for obtaining the 
quasiclassical eigenvalues of nonrelativistic Hamiltonians. It was 
demonstrated that the leading-order SWKB quantization condition in 
each and every case reproduces the exact energy eigenvalues for a 
class of solvable potentials. For these models, solutions can be 
written in terms of elementary functions.

Successes of the SWKB quantization rule have revived interest in the 
original WKB quantization condition. In several common applications 
the method gives very accurate results. Proofs of varying degrees of 
rigor have been advanced that demonstrate the exactness of the 
standard WKB quantization condition 
\cite{Fro},\cite{Sukh}-\cite{SuSe}.  The existing proofs of exactness 
of the WKB approximation are not entirely rigorous since the 
correction terms are only asymptotically valid, i.e., as $\hbar 
\rightarrow 0$ \cite{Krei}. Furthermore, in the cases when a modified 
WKB integral gives the exact eigenvalues, it is not even clear which 
``correction'' must be shown to be zero.

The standard lowest-order WKB prescription reproduces the exact 
energy levels for the harmonic oscillator in the Cartesian 
coordinates $x$, $y$, and $z$. But just this problem is correctly 
formulated in the framework of the quasiclassical approach; in the 
Cartesian coordinates, the Schr\"odinger's equation has the required 
canonical form and the generalized moments for each degree of freedom 
coincide with the corresponding classic moments. As for other 
coordinate systems, for example spherical, the WKB method does not 
reproduce the exact energy levels unless one supplements it with 
Langer-like correction terms.

For the central potential $V(r)$, the Schr\"odinger's equation can be 
written in the spherical coordinates as

\begin{eqnarray}
(-i\hbar)^2\left[\frac 1{r^2}\frac{\partial }{\partial r} \left(r^2
\frac{\partial }{\partial r}\right) + \frac 1{r^2\sin\,\theta}
\frac{\partial }{\partial\theta}\left(\sin\,\theta\frac{\partial}
{\partial \theta} \right) + \frac 1{r^2\sin^2\theta}\frac{\partial^2}
{\partial \varphi^2}\right]\psi(\vec r) =  \label{shr}  \\
2m[E-V(r)]\psi(\vec r).  \nonumber \end{eqnarray}
The standard solution of this equation is the following. If one 
substitutes $\psi(\vec r) = [U(r)/r]$ $[\Theta(\theta)/ 
\sqrt{\sin\theta}] \Phi(\varphi)$ into Eq. (\ref{shr}), one obtains 
(after separation) the following three reduced one-dimensional 
equations

\begin{equation}
\left[\hbar^2\frac{d^2}{dr^2} + 2m(E-V) - \frac{\vec L^2}{r^2}
\right]U(r) = 0,  \label{rad}   \end{equation}

\begin{equation}
\left(\hbar^2\frac{d^2}{d\theta^2} + \vec L^2 + \frac{\hbar^2}4 -
\frac{L_z^2 - \frac{\hbar^2}4}{\sin^2\theta}\right)
\Theta(\theta) = 0,   \label{tet}   \end{equation}

\begin{equation}
\left(\hbar^2\frac{d^2}{d\varphi^2} + L_z^2\right)\Phi(\varphi) = 0.
\label{phi}   \end{equation}
Exact solution of Eq. (\ref{tet}) gives, for the squared angular 
momentum $\vec L^2$, $\vec L^2=l(l+1)\hbar^2$. Application of the 
leading-order WKB quantization condition \cite{He},

\begin{equation}
\int_{x_1}^{x_2}\sqrt{p^2(x,E)}dx = \pi\hbar\left(n+\frac 12\right),
\ \ \ \ \ n=0,\,1,\,2,\,...,   \label{qc2}
\end{equation}
to the radial Eq. (\ref{rad}) does not reproduce the exact energy 
spectrum \{here in (\ref{qc2}) $x_1$ and $x_2$ are the classic 
turning points and $p^2(x,E)=2m[E-V(x)]$\}. The problem comes from 
the form of the centrifugal term, $l(l+1)\hbar^2/r^2$.

To overcome this problem in particular case of the Coulomb potential, 
a special techniques has been developed.  In order for the 
first-order WKB approximation to give the exact eigenvalues, the 
quantity $l(l+1)$ in Eq. (\ref{rad}) must be replaced by $(l+\frac 
12)^2$ \cite{Lang}. The reason for this modification (for the special 
case of the Coulomb potential) was pointed out by Langer ($1937$) 
\cite{Lang} from the Langer transformation $r = e^x$, $U(r) = 
e^{x/2}X(x)$. However, for other spherically symmetric potentials, in 
order to obtain the appropriate Langer-like correction terms, another 
special transformation of the wave function (w.f.) and its arguments 
is required.

There are several other related problems in the semiclassical 
consideration of the radial Schr\"odinger equation (\ref{rad}). (i) 
The WKB solution of the radial equation is irregular at $r\rightarrow 
0$, i.e. $R^{WKB}(r)\propto r^\lambda /\sqrt r$, $\lambda = 
\sqrt{l(l+1)}$, whereas the exact solution in this limit is 
$R(r)\propto r^l$. (ii) Equation (\ref{rad}) has no the centrifugal 
term when $l=0$, i.e. the radial problem has only one turning point 
and one can not use the WKB quantization condition (\ref{qc2}) 
derived for two-turning-point problems. However, solving the equation 
for $l>0$ by known exact methods one obtains energy eigenvalues for 
all $l$.  (iii) The WKB solution of equation (\ref{tet}) has 
analogous to the radial one, incorrect behavior at $\theta\rightarrow 
0$:  $\Theta^{WKB}(\theta) \propto\theta^\mu$, $\mu^2 = m^2 
-\hbar^2/4$, while the exact regular solution in this limit is 
$\Theta_l^m(\theta)$ $\propto \theta^{|m|}$. Angular eigenfunction 
$Y_{00}(\theta,\varphi) = const$, i.e. no nontrivial solution exists.

As practical use shows the standard leading-order WKB approximation 
{\em always} reproduces the exact spectrum for the solvable 
spherically symmetric potentials $V(r)$ if the centrifugal term in 
the radial Schr\"odinger's equation has the form $(l+\frac 
12)^2\hbar^2/r^2$.  As will be shown below the centrifugal term of 
such a form can be obtained from the WKB solution of equation 
(\ref{shr}) if separation of this three-dimensional equation has 
performed with the help of the correspondence principle. \\

\noindent {\bf 3. Separation of the Schr\"odinger's equation }\\

There are two essential features of the WKB method. First, the method 
was developed to solve the Schr\"odinger's equation in canonical form 
(without first derivatives). Second, in the quasiclassical method, 
the classic quantities such as classic momentum, classic action, 
phase, etc., are used. (For example, in the WKB quantization 
condition, the classic generalized momentum in the phase-space 
integral is used). For the harmonic oscillator in the Cartesian 
coordinates, the generalized moments in the original equation and 
moments obtained after separation coincide with the corresponding 
classic moments. Just for this problem, the standard WKB method (in 
one and multi-dimensional cases) reproduces the exact energy levels 
without any additional correction terms.

The generalized moments in Eqs. (\ref{rad})-(\ref{phi}) obtained from 
separation of Eq. (\ref{shr}) are different from the corresponding 
classic moments. As a result, the WKB method does not reproduce the 
exact energy levels for the spherically symmetric potentials (unless 
one supplements it with Langer-like correction terms). The reason is 
the form of the squared angular momentum, $\vec L^2=l(l+1)\hbar^2$, 
which is obtained from solution of the equation (3). In the WKB 
method, in order to reproduce the exact energy spectrum, the term 
$(l+\frac 12)^2\hbar^2$ should be used in the centrifugal term. This 
term, $M^2=\vec L^2+\hbar/4$, is in the angular equation (\ref{tet}), 
but is not in the radial equation (\ref{rad}).

Let us show that the term $(l+\frac 12)^2\hbar^2$ can be obtained 
from the quasiclassical solution of the reduced Schr\"odinger's 
equation.  For this, exclude in Eq. (\ref{shr}) the first derivatives 
that can be easily done with the help of the following operator 
identity:

\begin{equation}
\frac d{dx}g(x)\frac d{dx} \equiv \left[\sqrt{g(x)}\frac{d^2}{dx^2} -
\frac{d^2}{dx^2}\sqrt{g(x)}\right]\sqrt{g(x)}.  \label{opid}
\end{equation}
Then, after dividing by $\tilde\psi(\vec r) =\tilde R(r)
\tilde\Theta(\theta)\tilde\Phi(\varphi)$, where $\tilde R(r) =
rR(r)$, $\tilde\Theta(\theta) = \sqrt{\sin(\theta)}\Theta(\theta)$,
$\tilde\Phi(\varphi) = \Phi(\varphi)$, we obtain the equation
\begin{eqnarray}
-\hbar^2\frac{\tilde R_{rr}''}{\tilde R} +
\frac 1{r^2}\left(-\hbar^2\frac{\tilde\Theta_{\theta\theta}''}
{\tilde\Theta} - \frac{\hbar^2}4\right) +
\frac 1{r^2\sin^2\theta}\left(-\hbar^2\frac{\tilde
\Phi_{\varphi\varphi}''}{\tilde\Phi} -
\frac{\hbar^2}4\right) =   \label{rede}  \\
 2m[E - V(r)].   \nonumber
\end{eqnarray}
Introducing the notations,

\begin{equation}
\left(\frac{\partial S_0}{\partial r}\right)^2 =
-\hbar^2\frac{\tilde R_{rr}''}{\tilde R},   \label{sr}
\end{equation}

\begin{equation}
\left(\frac{\partial S_0}{\partial\theta}\right)^2 =
-\hbar^2\frac{\tilde\Theta_{\theta\theta}''}{\tilde\Theta} -
\frac{\hbar^2}4,  \label{st}
\end{equation}

\begin{equation}
\left(\frac{\partial S_0}{\partial\varphi}\right)^2 =
-\hbar^2\frac{\tilde\Phi_{\varphi\varphi}''}{\tilde\Phi} -
\frac{\hbar^2}4,  \label{sp}
\end{equation}
we can write Eq. (\ref{rede}) in the form of the classic 
Hamilton-Jacobi equation,

\begin{equation}
\left(\frac{\partial S_0}{\partial r}\right)^2 + \frac
1{r^2}\left(\frac{ \partial S_0}{\partial\theta }\right)^2 +
\frac 1{r^2\sin^2\theta}\left( \frac{\partial S_0}{\partial\varphi}
\right)^2 = 2m\left[E-V(r)\right],  \label{HJ}
\end{equation}
where $S_0=S_0(\vec r,E)$ is the classic action of the system.

Using the correspondence principle, we see, from Eq. (\ref{HJ}), that 
Eqs. (\ref{sr})-(\ref{sp}) are the squared generalized moments 
expressed via the quantum-mechanical quantities. Now, let us separate 
equation (\ref{HJ}). Then, taking into account Eqs.  
(\ref{sr})-(\ref{sp}), we obtain the following system of the 
second-order differential equations in canonical form

\begin{equation}
\left(-i\hbar\frac d{dr}\right)^2\tilde R =
\left[2m(E-V) - \frac{\vec M^2}{r^2}\right]\tilde R,  \label{rra}
\end{equation}

\begin{equation}
\left[\left(-i\hbar\frac d{d\theta}\right)^2 - \left(\frac\hbar
2\right)^2\right]\tilde\Theta(\theta) = \left(\vec M^2 -
\frac{M_z^2}{\sin^2\theta}\right)\tilde\Theta(\theta),   \label{rth}
\end{equation}

\begin{equation}
\left[\left(-i\hbar\frac d{d\varphi}\right)^2 - \left(\frac\hbar
2\right)^2\right]\tilde\Phi(\varphi) = M_z^2\tilde\Phi(\varphi),
\label{rph} \end{equation}
where $\vec M^2$ and $M_z^2$ are the constants of separation and, at 
the same time, integrals of motion.

Equations (\ref{rra})-(\ref{rph}) have the quantum-mechanical form 
$\hat f\psi = f\psi$, where $f$ is the physical quantity (the squared 
generalized momentum) and $\hat f$ is the corresponding operator. We 
see that the term $\hbar^2/4$ in the left-hand side of the equations 
is related to the squared angular momentum operator. This term 
disappears in the leading $\hbar$ approximation \cite{SeS} and the 
equations (\ref{rra})-(\ref{rph}) can be written in the general form 
as

\begin{equation}
\left(-i\hbar\frac d{dq}\right)^2\psi(q) = [\lambda^2-U(q)]\psi(q).
\label{gen}  \end{equation}
The squared generalized moments in the right-hand sides are the same 
as the classic ones. The use of these moments in the WKB quantization 
condition and WKB solution yields the exact energy spectra for the 
central-field potentials and does not result in the difficulties of 
the WKB method mentioned above.\\

\noindent {\bf 4. Solution of the Schr\"odinger's equation }\\

In this section we consider quasiclassical solution of the 
Schr\"odinger's equation for the spherically symmetric potentials. In 
order for the approach we consider here to be self-consistent we have 
to solve each equation obtained after separation by the same method, 
i.e. the WKB method. The quasiclassical  method is general enough and 
the WKB formulas can be written differently, i.e. on the real axis 
\cite{He} and in the complex plane \cite{Fro}. Most general form of 
the WKB solution and quantization condition can be written in the 
complex plane.\\

{\bf The WKB quantization in the complex plane.} The WKB method is 
usually used to solve one-dimensional two turning point problems. 
Within the framework of the WKB method the solvable potentials mean 
those potentials for which the eigenvalue problem has two turning 
points. However the WKB method can be used to solve problems with 
more then two turning points.  In this case formulation in the 
complex plane is the most appropriate.

Consider Eqs. (\ref{rra})-(\ref{rph}) in the framework of the 
quasiclassical method. Solution of each of these equations [in the 
general form Eq. (\ref{gen})] we search in the form \cite{Fro}

\begin{equation}
\psi_\lambda(z) = A\,\exp\left[\frac i{\hbar}S(z,\lambda)\right],
\label{psiS}   \end{equation}
where $A$ is the arbitrary constant. The function $S(z,\lambda)$ is 
written as the expansion in powers of $\hbar$, $S(z,\lambda)=$ 
$S_0(z,\lambda) + \hbar S_1(z,\lambda) +\frac 12\hbar^2S_2(z,\lambda) 
+\dots$. In the leading $\hbar$ approximation the WKB solution of 
Eqs. (\ref{rra})-(\ref{rph}) can be written in the form

\begin{equation}
\psi^{WKB}(z) = \frac A{\sqrt{p(z,\lambda)}} \exp\left[\pm\frac
i{\hbar}\int_{z_0}^z\sqrt{p^2(z,\lambda)}dz\right].  \label{wfWKB}
\end{equation}

In quantum mechanics, quantum numbers are determined as number of 
zeroes of the w.f. in the physical region. In the complex plane, the 
number of zeroes $N$ of a function $y(z)$ inside the contour $C$ is 
defined by the argument's principle \cite{Kor,Wen}. For the w.f.  
$\psi_\lambda(z)$, according to this principle we have

\begin{equation}
\oint_C \frac{\psi_\lambda^\prime(z)}{\psi_\lambda(z)}dz = 2\pi iN,
\label{argp}  \end{equation}
where $\psi_\lambda^\prime(z)$ is the derivative of the function 
$\psi_\lambda(z) $ over the variable $z$ [see Ref. \cite{FF} for more 
information about the condition (\ref{argp})]. Contour $C$ is chosen 
such that it includes cuts (therefore, zeroes of the w.f.) between 
the turning points where $p^2(z,\lambda)=\lambda^2 - U(z)>0$.

Substitution of Eq. (\ref{wfWKB}) into (\ref{argp}) results in the
quantization condition

\begin{equation} \oint\sqrt{p^2(z,\lambda)}dz +
i\frac\hbar 2\oint\frac{p^\prime(z,\lambda)}{p(z,\lambda)}dz =
2\pi\hbar N.  \label{gqc}   \end{equation}
In the case, when $p(z,\lambda)$ is a smooth function of the spatial 
variable and the equation $\lambda^2-U(z)=0$ has two roots (turning 
points), the quantization condition (\ref{gqc}) takes the form

\begin{equation}
\oint\sqrt{p^2(z,\lambda)}dz = 2\pi\hbar\left(N+\frac 12\right).
\label{gqc2}  \end{equation}
In particular, for $p^2(z,\lambda)=\lambda^2$, the quantization 
condition is

\begin{equation}
\oint\sqrt{p^2(z,\lambda)}dz = 2\pi\hbar N.  \label{qcc}
\end{equation}
In the next section we solve the three-dimensional Schr\"odinger 
equation for several spherically symmetric potentials by the method 
under consideration.  \\

{\bf A. The angular momentum eigenvalues}.
Equations (\ref{rth}) and (\ref{rph}) determine the squared angular 
momentum eigenvalues, $\vec M^2$, and its projection, $M_z$, 
respectively. The quantization condition (\ref{qcc}) appropriate to 
the angular equation (\ref{rph}),

\begin{equation}
\oint M_zd\varphi = 2\pi\hbar m,     \label{qcm}   \end{equation}
gives $M_z = \hbar m$, $m = 0,1,2,...$ The corresponding
quasiclassical solution is

\begin{equation}
\tilde\Phi_m(\varphi) = C_1\,e^{im\varphi} + C_2\,e^{-im\varphi},
\label{sph} \end{equation}
where $C_1$ and $C_2$ are the arbitrary constants.

The quantization condition (\ref{gqc2}) appropriate to Eq.
(\ref{rth}) is

\begin{equation}
I = \oint_C\sqrt{\vec M^2 - \frac{M_z^2}{\sin^2\theta}}d\theta =
2\pi\hbar\left(n_{\theta} +\frac 12\right), \ \ \ n_{\theta} =
0,1,2,...  \label{Ith}
\end{equation}
To calculate the integral (\ref{Ith}) (as other hereafter) we use the 
method of stereographic projection. This means that, instead of 
integration about a contour $C$ enclosing the classical turning 
points, we exclude the singularities outside the contour $C$, i.e., 
at $\theta = 0$ and $\infty $ in this particular case. Excluding 
these infinities we have, for the integral (\ref{Ith}), $I = I_0 + 
I_{\infty}$. Integral $I_0 = -2\pi M_z$, and $I_{\infty}$ is 
calculated with the help of the replacement $z=e^{i\theta}$ that 
gives $I_{\infty} = 2\pi\sqrt{\vec M^2}\equiv2\pi M$. Therefore, $I = 
2\pi(M - M_z)$ and we obtain, for the squared angular momentum 
eigenvalues,

\begin{equation}  \vec M^2 = \left(l+\frac 12\right)^2\hbar^2,
\label{M2} \end{equation}
where $l=n_{\theta}+m$. Thus the quasiclassical solution of the 
Schr\"odinger's equation results in the squared angular momentum 
eigenvalues (\ref{M2}). This means the centrifugal term in the radial 
Eq. (\ref{rra}) has the form $(l+\frac 12)^2\hbar^2/r^2$ for {\em 
all} spherically symmetric potentials.

As known the WKB solution $\Theta^{WKB}(\theta)$ of the equation 
(\ref{tet}) has incorrect asymptotes at $\theta\rightarrow 0$ and 
$\pi$. At the same time, the WKB solution of Eq. (\ref{rth}), which 
corresponds to the eigenvalues (\ref{M2}), has the correct asymptotic 
behavior at these points for all $l$. So far, as the generalized 
momentum $p(\theta)\simeq\frac{\mid m\mid}\theta$ at 
$\theta\rightarrow 0$, this gives, for the WKB solution in the 
representation of the wave function $\psi(\vec r)$, 
$\Theta_l^m(\theta) = \tilde\Theta^{WKB} (\theta)/ 
\sqrt{\sin\,\theta} \propto\theta^{|m|}$ which corresponds to the 
behavior of the known exact solution $Y_{lm}(\theta ,\varphi)$ at 
$\theta\rightarrow 0$.

In the classically allowed region, where $p^2(\theta,M)=\vec 
M^2-M_z^2/\sin^2\theta >0$, the leading-order WKB solution of Eq. 
(\ref{rth}) is

\begin{equation}
\tilde{\Theta}^{WKB}(\theta) = \frac B{\sqrt{p(\theta,M)}}
\cos\left[\int_{\theta_1}^{\theta}p(\theta,M)d\theta -\frac\pi 4 \right].
\label{solth}  \end{equation}
The normalized quasiclassical solution far from the turning points, 
where $p(\theta,M)\simeq (l+\frac 12)\hbar$, can be written in 
elementary functions as

\begin{equation}
\tilde{\Theta}_l^m(\theta) = \sqrt{\frac{2l+1}{\pi(l-m +\frac 12)}}
\cos\left[\left(l+\frac 12\right)\theta +
\frac{\pi}2(l-m) \right],  \label{Thet}
\end{equation}
where we have took into account that the phase-space integral at the 
classic turning point $\theta_1$ is $\chi(\theta_1)=$ $-\frac\pi 
2(n_\theta +\frac 12)$ and $\chi(\theta_2)=$ $\frac\pi 2(n_\theta 
+\frac 12)$ at $\theta=\theta_2$. We see that the eigenfunctions 
(\ref{Thet}) are either symmetric or antisymmetric.  The 
corresponding WKB solution, $\tilde Y_{lm}^{WKB}(\theta,\varphi)=$ 
$\tilde{\Theta}_l^m(\theta)\tilde\Phi_m(\varphi)$, where the 
normalized eigenfunction $\tilde\Phi_m(\varphi)=\frac 1{\sqrt{2\pi}}$ 
$e^{\pm im\varphi}$, in the representation of the w.f.  
$\tilde\psi(\vec r)$ is

\begin{equation} \tilde Y_{lm}^{WKB}(\theta,\varphi) =
\frac 1{\pi}\sqrt{\frac{l+\frac 12}{l - m + \frac 12}}
\cos\left[\left(l+\frac 12\right)\theta + \frac{\pi}2(l - m)\right]
e^{\pm im\varphi}.   \label{Ythet}
\end{equation}

Remind some results concerning the semiclassical approach in quantum 
mechanics. The general form of the semiclassical description of 
quantum-mechanical systems have been considered in Ref. \cite{Mi}.  
It was shown that the semiclassical description resulting from Weyl's 
association of operators to functions is identical with the quantum 
description and no information need to be lost in going from one to 
the another. What is more "the semiclassical description is more 
general than quantum mechanical description..."  \cite{Mi}. The 
semiclassical approach merely becomes a different representation of 
the same algebra as that of the quantum mechanical system, and then 
the expectation values, dispersions, and dynamics of both become 
identical.

One of the fundamental features of quantum mechanical systems is 
nonzero minimal energy which corresponds to quantum oscillations. The 
corresponding w.f. has no zeroes in the physical region. Typical 
example is the harmonic oscillator.

Eigenvalues of the one-dimensional harmonic oscillator are $E_n = 
\hbar\omega(n+\frac 12)$, i.e. the energy of zeroth oscillations $E_0 
= \frac 12 \hbar\omega$. In three-dimensional case, in the Cartesian 
coordinates, the eigenvalues of the oscillator are $E_n = 
\hbar\omega(n_x+n_y+n_z+\frac 32)$ \cite{Flu}, i.e. each degree of 
freedom contributes to the energy of the ground state, $E_0 = E_{0,x} 
+ E_{0,y} + E_{0,z} = $ $\frac 32\hbar\omega$. Energy of the ground 
state should not depend on coordinate system. This means that, in the 
spherical coordinates, each degree of freedom (radial and angular) 
should contribute to the energy of zeroth oscillations.  In many 
applications and physical models a nonzero minimal angular momentum 
$M_0$ is introduced (phenomenologically) in order to obtain 
physically meaningful result (see, for instance, Ref. \cite{Iwa}).  
However, the existence of $M_0$ follows from the quasiclassical 
solution of Eq. (\ref{rth}) \cite{SeS,SeF}.

Consider the WKB eigenfunction (\ref{Ythet}) for the ground state.  
Setting in (\ref{Ythet}) $m=0$ and $l=0$, we obtain the nontrivial 
solution in the form of a standing half-wave [remind that the 
spherical function $Y_{00}(\theta,\varphi)=const$],

\begin{equation}
\tilde Y_{00}^{WKB}(\theta,\varphi) = \frac 1\pi\cos\frac\theta 2.
\label{Y0}   \end{equation}
The corresponding eigenvalue is

\begin{equation}  M_0 = \frac{\hbar}2.  \label{M0}
\end{equation}
The eigenvalue (\ref{M0}) contributes to the energy of zeroth 
oscillations. This means that (\ref{Y0}) can be considered as 
solution, which describes the quantum fluctuations of the angular 
momentum. Note, that the eigenfunction of the ground state, $\tilde 
Y_{00}^{WKB}(\theta,\varphi)$, is symmetric. Below, we solve the 
radial equation (\ref{rra}) for some spherically symmetric potentials 
and show the contribution of the eigenvalue $M_0$ to the energy of 
the ground state.\\

{\bf B. The Coulomb problem $V(r)=-\frac{\alpha}r$}.
The WKB quantization condition (19) appropriate to the radial 
equation (\ref{rad}) with the Coulomb potential does not reproduce 
the exact energy levels unless one supplements it with Langer-like 
correction terms. Another problem is that the radial Schr\"odinger's 
equation (\ref{rad}) has no the centrifugal term at $l=0$ and one can 
not use the WKB quantization condition (derived for two-turning-point 
problems) to calculate the energy of the ground state directly from 
this equation.  We do not run into such a problem in case of Eqs.  
(\ref{rra})-(\ref{rph}). As follows from the above consideration, the 
centrifugal term $(l+\frac 12)^2\hbar^2/r^2$ is the same for all 
spherically symmetric potentials and the WKB method reproduces the 
exact energy spectrum for all $l$ and $n_r$.

For the radial equation (\ref{rra}) with the Coulomb potential, the 
WKB quantization condition (\ref{gqc2}) is

\begin{equation}  I =\oint_C\sqrt{2mE +\frac{2m\alpha}r - \frac{\vec
M^2}{r^2}}dr = 2\pi\hbar\left(n_r+\frac 12\right),   \label{Icou}
\end{equation}
where the integral is taken about a contour $C$ inclosing the turning 
points $r_1$ and $r_2$. Using the method of stereographic projection, 
we should exclude the singularities outside the contour $C$, i.e. at 
$r=0$ and $\infty$. Excluding these infinities we have, for the 
integral (\ref{Icou}), $I = I_0 + I_{\infty}$, where $I_0 = 2\pi 
i\sqrt{-\vec M^2}\equiv -2\pi M$ and $I_{\infty} = 2\pi i\alpha m/ 
\sqrt{2mE}$. The sequential simple calculations result in the exact 
energy spectrum

\begin{equation}
 E_n = -\frac{\alpha^2m}{2[(n_r+\frac 12)\hbar + M]^2}. \label{Ecou}
\end{equation}

For the energy of zeroth oscillations we have, from Eq. (\ref{Ecou}), 
$E_0 = -\frac 12\alpha^2 m(\frac{\hbar}2 + M_0)^{-2}$, that 
apparently shows the contribution of the quantum fluctuations of the 
angular momentum (see Eq. (\ref{M0})) into the energy of the ground 
state $E_0$ \cite{SeF}. The radial quasiclassical eigenfunctions, 
$\tilde R_n^{WKB}(r)$, inside the classical region $[r_1,r_2]$ far 
from the turning points $r_1$ and $r_2$ are written in elementary 
functions in the form of a standing wave \cite{SeS},

\begin{equation}
\tilde R_n(r) = A\cos\left(\frac 1\hbar p_nr +\frac\pi 2n_r\right),
\label{Rn}  \end{equation}
where we have took into account that the phase-space integral 
(\ref{Icou}) at the classic turning point $r_1$ is $\chi(r_1)=$ 
$-\frac\pi 2(n_r +\frac 12)$. Here $A$ is the normalization constant 
and $p_n$ is the eigenmomentum expressed via the energy eigenvalue 
$E_n$, $p_n = \sqrt{2m|E_n|}$. The eigenfunctions (\ref{Rn}) are 
either symmetric or antisymmetric. \\

{\bf C. The three-dimensional harmonic oscillator
$V(r)=\frac 12 m\omega^2r^2$}.
The three-dimensional harmonic oscillator is another classic example 
of the exactly solvable problems in quantum mechanics. The problem 
has $4$ turning points, $r_1$, $r_2$, $r_3$, and $r_4$, but only two 
of them, $r_3$ and $r_4$, lie in the physical region $r>0$.

The problem is usually solved with the help of the replacement 
$x=r^2$ which reduces the problem to the $2$-turning-point ($2$TP) 
one. But this problem can be solved as the $4$TP problem in the 
complex plane. Because of importance of the oscillator potential in 
many applications and with the purpose of further development of the 
WKB method, we shall solve the problem by two methods, on the real 
axis as $2$TP problem and then in the complex plane as $4$TP problem.

Consider first the physical region $r>0$, where the problem has two
turning points. The leading-order WKB quantization condition
(\ref{gqc2}) then is

\begin{equation}
I =\int_{r_3}^{r_4}\sqrt{2mE - (m\omega r)^2 -
\frac{\vec M^2}{r^2}}dr = \pi\hbar\left(n_r+\frac 12\right),
\label{Io}  \end{equation}
where $n_r$ is the number of zeroes of the w.f.  between the classic 
turning points $r_3$ and $r_4$. Integral (\ref{Io}) is reduced to the 
above case of the Coulomb potential with the help of the replacement 
$z=r^2$. Integration result is $I=\pi(E/\omega -M)/2$ and we obtain, 
for the energy eigenvalues,

\begin{equation}
E_n = \omega\left[2\hbar\left(n_r + \frac 12\right) + M\right].
\label{Eosc}    \end{equation}
So far, as $M = (l+\frac 12)\hbar$, we obtain the exact energy
spectrum for the isotropic oscillator. Energy of the ground state is 
$E_0 =\omega(\hbar + M_0)$, where $M_0$ is the contribution of 
quantum fluctuations of the angular momentum.

Emphasize the following in this solution. The $4$TP problem has been 
solved as the $2$TP problem; we have applied the $2$TP quantization 
condition (\ref{gqc2}) to the $4$TP problem that is not quite 
correct. We have obtained the correct result because the potential is 
symmetric and the replacement $x=r^2$ reduces the problem to the 
$2$TP problem, i.e.``reflects'' the negative region $r<0$ (and zeroes 
of the w.f.) into the positive region. A more correct approach to 
solve the problem would be a $4$TP quantization condition.  
Fortunately, the WKB method in the complex plane allows to solve this 
problem as the $4$TP problem.

In the complex plane, the problem has two cuts, between turning 
points $r_1$, $r_2$ and $r_3$, $r_4$. To apply residue theory for the 
phase space integral we need to take into account all zeroes of the 
w.f. in the complex plane, i.e. the contour $C$ has to include both 
cuts. The quantization condition (\ref{gqc}) in this case takes the 
form

\begin{eqnarray}
\oint_C\left[p(r,E) +
i\frac\hbar 2\frac{p^\prime(r,E)}{p(r,E)}\right]dr
\equiv   \label{sumC}  \\   \oint_{C_1}\left[p(r,E)+i\frac\hbar
2\frac{p^\prime(r,E)}{p(r,E)}\right]dr +
\oint_{C_2}\left[p(r,E)+i\frac\hbar
2\frac{p^\prime(r,E)}{p(r,E)}\right]dr = 2\pi\hbar N,  \nonumber
\end{eqnarray}
where $p^2(r,E)=2mE -(m\omega r)^2 -\vec M^2/r^2$, and $C_1$ and
$C_2$ are the contours about the cuts at $r<0$ and $r>0$, respectively.
The number $N=n_{r<0}+n_{r>0}$, where $n_{r<0}$ and $n_{r>0}$ are
the numbers of zeroes of the w.f. at $r<0$ and  $r>0$, respectively.
For the harmonic oscillator, because of symmetricity of the potential
we have $n_{r<0}=n_{r>0}=n_r$, i.e. the total number of zeroes is
$N=2n_r$.

Therefore, the quantization condition (\ref{sumC}) for the $4$TP 
problem takes the form,

\begin{equation}
\oint_Cp(r,E)dr = \oint_{C_1}p(r,E)dr + \oint_{C_2}p(r,E)dr =
2\pi\hbar k\left(n_r+\frac 12\right),  \label{oink}
\end{equation}
where $k=2$ is the number of cuts. We can write the $4$TP 
quantization condition in this form because the effective potential 
is infinite at $r=0$. In case if the potential is finite in the whole 
region, the quantization condition will be more complicate 
\cite{SuSe}.

The condition (\ref{oink}) is in agreement with the Maslov's theory. 
This means that the right-hand side of the equation (\ref{oink}) can 
be written in the form

\begin{equation}
2\pi\hbar k\left(n_r+\frac 12\right) =
2\pi\hbar\left(N +\frac\mu 4\right),  \label{qmas}
\end{equation}
where $\mu=2k$ is the Maslov's index, i.e. number of reflections of 
the w.f. on the walls of the potential.

In the general case of the potential which is infinite between cuts, 
the $2k$ turning point quantization condition is

\begin{equation}
\oint_C p(z,E)dz =
2\pi\hbar\sum_{i=1}^k \left(n+\frac 12\right)_i \equiv
2\pi\hbar\left(N+\frac\mu 4\right),   \label{genC}
\end{equation}
where $N=kn_i$ is the total number of zeroes of the w.f. on the $k$ 
cuts. On the real axis, for the $2k$ TP problem, the quantization 
condition (\ref{genC}) has the form

\begin{equation}
\sum_{i=1}^k\int_{x_{1i}}^{x_{2i}}\sqrt{p^2(z,E)}dz =
\pi\hbar\left(N+\frac\mu 4\right).   \label{suxi}
\end{equation}

Because the harmonic oscillator potential is symmetric, integrals in 
Eq. (\ref{oink}) are identical, i.e. the quantization condition 
(\ref{oink}) is equivalent to the $2$TP quantization condition 
(\ref{gqc2}). The phase-space integral can be easily calculated in 
the complex plane. For this we have to exclude the singularities at 
$r=0$ and $\infty$ outside the contour $C$. Excluding these 
infinities we have, for the integral (\ref{genC}) with the isotropic 
potential, $I = I_0 + I_{\infty}$, where $I_0 = -2\pi M$ and integral 
$I_{\infty}$ is calculated with the help of the replacement $r=1/z$, 
$I_\infty=2\pi E/\omega$, i.e. we again obtain the exact result 
(\ref{Eosc}) for $E_n$.

Consider Eq. (\ref{Eosc}) at $n_r=0$ and $l=0$, i.e. the energy of 
the ground state. We have $E_0 = \omega(\hbar + M_0)$, where $M_0 
=\hbar /2$ is the contribution of the quantum fluctuations of the 
angular momentum into the energy of the ground state $E_0$. The 
radial quasiclassical eigenfunctions, $\tilde R_n^{WKB}(r)$, in the 
region of the classical motion far from the turning points are 
written analogously to the above case in the form of a standing wave 
[see Eq. (\ref{Rn})] .\\

{\bf D. The Hulth\'en potential $V(r)=-V_0e^{-r/r_0}/
(1-e^{-r/r_0})$}.
The Hulth\'en potential is of a special interest in atomic and 
molecular physics. The potential is known as an ``insoluble'' by the 
standard WKB method potentials, unless one supplements it with 
Langer-like corrections. The radial problem for this potential is 
usually considered at $l=0$. However, in the approach under 
consideration, the quasiclassical method results in the nonzero 
centrifugal term at $l=0$ and allows to obtain the analytic result 
for all $l$.

The leading-order WKB quantization condition (\ref{gqc2}) for the
Hulth\'en potential is

\begin{equation}
I=\oint\sqrt{2m\left( E + V_0\frac{e^{-r/r_0}}
{1-e^{-r/r_0}}\right) -\frac{\vec M^2}{r^2}}dr = 2\pi
\hbar\left(n_r+\frac 12\right).   \label{Ihul}
\end{equation}
In the region $r>0$, this problem has two turning points $r_1$ and 
$r_2$. The phase-space integral (\ref{Ihul}) is calculated 
analogously to the above case. Introducing the new variable $\rho = 
r/r_0$, we calculate the contour integral in the complex plane, where 
the contour $C$ encloses the classical turning points $\rho_1$ and 
$\rho_2$. Using the method of stereographic projection, we should 
exclude the infinities at $r=0$ and $\infty$ outside the contour $C$. 
Excluding these infinities we have, for the integral (\ref{Ihul}), 
$I=I_0+I_\infty$, where $I_0=-2\pi M$ and $I_\infty$ is calculated 
with the help of the replacement $z=e^\rho -1$ \cite{SeS},

\begin{eqnarray} I = \oint \sqrt{2mr_0^2\left(E + V_0
\frac{e^{-\rho }}{1-e^{-\rho }}\right) - \frac{\vec M^2}
{\rho ^2}}d\rho =  \label{Ihul1}  \\  -2\pi M + 2\pi
r_0\sqrt{-2m}\left[ -\sqrt{-E} + \sqrt{-E+V_0}\right]. \nonumber
\end{eqnarray}

Substituting the integration result into Eq. (\ref{Ihul}), we 
immediately get the exact energy spectrum

\begin{equation} E_n=-\frac 1{8mr_0^2}\left(\frac{2mV_0r_0^2}N -
N\right)^2.  \label{Ehul}  \end{equation}
where $N = (n_r+\frac 12)\hbar + M$ is the principal quantum number.  
Setting in (\ref{Ehul}) $M=0$, we arrive at the energy eigenvalues 
obtained from known exact solution of the Schr\"odinger's equation at 
$l=0$. However, in our case $M_{min}\equiv M_0 = \hbar/2$ at $l=0$ 
and the principal quantum number is $N = (n_r+\frac 12)\hbar + M_0$.  
As in the previous examples, this apparently shows the contribution 
of the quantum fluctuations of the angular momentum into the energy 
of the ground state, $E_0$.\\

{\bf E. The Morse potential $V(r) = V_0[e^{-2\alpha (r/r_0-1)}-2$
$e^{-\alpha (r/r_0-1)}]$}.
The Morse potential is usually considered as one-dimensional problem 
at $l=0$. In this case the problem has two turning points (note that 
the left turning point, $r_1$, is negative) and can be solved 
exactly.  In the general case, for $l>0$, we have an ``insoluble'' 
$4$TP problem.

For this potential, let us consider, first, the radial Schr\"odinger 
equation (\ref{rad}), which does not contain the centrifugal term at 
$l=0$,

\begin{equation}   \left(-i\hbar\frac d{dr}\right)^2U(r) =
2m\left[E-V_0 e^{-2\alpha(r-r_0)/r_0}+2V_0e^{-\alpha (r-r_0)/r_0}
\right]U(r).   \label{shMor}
\end{equation}
The first-order WKB quantization condition (\ref{qc2}) appropriate to 
this equation is

\begin{equation}  \int_{r_1}^{r_2}\sqrt{2m[E -
V_0e^{-2\alpha (r-r_0)/r_0} + 2V_0e^{-\alpha (r-r_0)/r_0}]}dr =
\pi\hbar\left(n_r+\frac 12\right).   \label{IMor}
\end{equation}
Introducing a variable $x=e^{-\alpha (r-r_0)/r_0}$, we reduce the 
phase-space integral to the well known one. Sequential simple 
calculations result in the exact energy eigenvalues

\begin{equation} E_n = -V_0\left[1-\frac{\alpha\hbar(n_r +\frac 12)}
{r_0\sqrt{2mV_0}} \right]^2.   \label{EMor}
\end{equation}

Now, let us deal with Eq. (\ref{rra}) for this potential, which 
contains the non-vanishing centrifugal term, $\hbar^2/4r^2$, at 
$l=0$. In this case we have an ``insoluble'' $4$TP problem. In the 
complex plane, the problem has two cuts ($k=2$), at $r<0$ and $r>0$, 
therefore, we apply the $4$TP quantization condition (\ref{oink}),

\begin{equation}
I = \oint_C\sqrt{2m\left[ E-V_0e^{-2\alpha (r-r_0)/r_0} +
2V_0e^{-\alpha (r-r_0)/r_0}\right] -\frac{\vec M^2}{r^2}}dr =
4\pi\hbar \left(n_r +\frac 12\right),   \label{IMoM}
\end{equation}
where the contour $C$ encloses the two cuts, but does not enclose the 
point $r=0$. To calculate this integral introduce the variable $\rho 
= r/r_0$. Using the method of stereographic projection, we should 
exclude the singularities outside the contour $C$, i.e. at $r=0$ and 
$\infty$. Excluding these infinities we have, for the integral 
(\ref{IMoM}) \cite{SeS},

\begin{equation}
I = -2\pi M -\frac{2\pi r_0}{\alpha}\left(\sqrt{-2mE} -
\sqrt{2mV_0}\right),    \label{IMocl} \end{equation}
and for the energy eigenvalues this gives

\begin{equation}
E_n = -V_0\left[1-\alpha\frac{2\hbar(n_r+\frac 12) +
M}{r_0\sqrt{2mV_0}} \right]^2.  \label{EMorM}
\end{equation}

Setting in (\ref{EMorM}) $l=0$, we obtain,

\begin{equation}
E_n = -V_0\left[1-\frac{\alpha[2\hbar(n_r+\frac 12)+M_0}
{r_0\sqrt{2mV_0}} \right]^2. \label{EMoM0}
\end{equation}
Equation (\ref{EMoM0}) for $E_n$ is different from the expression 
(\ref{EMor}) obtained from solution of Eq. (\ref{rad}) for the Morse 
potential at $l=0$. This difference is caused by the nonzero 
centrifugal term $\hbar^2/4r^2$ in the radial equation (\ref{rra}) at 
$l=0$. Thus we obtain two results for the Morse potential by the WKB 
method: the known exact eigenvalues (\ref{EMor}) obtained from 
solution of Eq.  (\ref{rad}) at $l=0$ and another result 
(\ref{EMorM}) obtained from solution of Eq. (\ref{rra}) for all 
$l$.\\

{\bf F. The potential $V(r) = kr +\frac 12\omega^2r^2$}.
This linear plus isotropic potential is one of the interest in 
particle physics. The potential has four turning points and can not 
be reduced to a hypergeometric function by a suitable transformation.  
This multi-turning-point problem is ``insoluble'', also, by the 
standard WKB method. However, the WKB method in the complex plane 
allows easily to solve this $4$TP problem [see, for instance, Refs. 
\cite{SeR,KrSe}].

The problem has two cuts ($k=2$) in the complex plane between turning 
points $r_1$, $r_2$ and $r_3$, $r_4$. Quantization condition 
(\ref{genC}), for this problem, is

\begin{equation}  I =\oint_C\sqrt{2mE - 2mkr - (m\omega r)^2 -
\frac{\vec M^2}{r^2}}dr = 4\pi\hbar\left(n_r+\frac 12\right),
\label{Ilq} \end{equation}
where the contour $C$ includes both cuts, but includes no the point 
$r=0$. To calculate the integral (\ref{Ilq}) we exclude the 
infinities at $r=0$ and $\infty$ outside the contour $C$. Excluding 
these infinities we have, for the integral (\ref{Ilq}),

\begin{equation}  I = 2\pi\left[\frac E\omega + \frac 1{2m\omega}
\left(\frac k{\omega}\right)^2 - M\right] =
4\pi\hbar\left(n_r +\frac 12\right),  \label{Ilqc}
\end{equation}
or, for the energy eigenvalues, this gives

\begin{equation}
E_n = \omega\left[2\hbar\left(n_r + \frac 12\right) + M\right] -
\frac 1{2m}\left(\frac k\omega\right)^2.  \label{Elq}
\end{equation}

Energy eigenvalues (\ref{Elq}) are similar to the harmonic oscillator 
ones but shifted by the constant value $-\frac 1{2m}(k/\omega)^2$.  
Putting in Eq. (\ref{Elq}) $k=0$, we arrive to the eigenvalues 
(\ref{Eosc}) for the isotropic oscillator.

Analogously one can obtain energy eigenvalues for other spherically 
symmetric potentials. The standard leading-order WKB approximation 
appropriate to the wave equation (\ref{rra}) yields the exact energy 
eigenvalues for known solvable potentials and ``insoluble'' ones with 
more than two turning points. This is possible because the 
centrifugal term of the required form, $(l+\frac 12)^2\hbar^2/r^2$, 
has obtained in a natural way from solution of the angular equation 
(\ref{rth}) with the use of the same WKB method; this term is the 
same for all central-field potentials. \\ \newpage

\noindent {\bf 5. Conclusion } \\

Conventional approach to solve the Schr\"odinger's equation is to 
reduce the original equation to a hypergeometric form or some special 
function by a suitable transformation. In each and every case, one 
needs to find, first, a special transformation for the wave function 
and its arguments to reduce the original equation to some known 
equation. There is another way to solve the Schr\"odinger's equation 
which is simple, general for all types of problems in quantum 
mechanics, and a very efficient to solve not only two- but 
multi-turning point problems.

Almost together with quantum mechanics an appropriate method to solve 
the wave equation has been developed known mainly as the WKB 
approximation. This method is general for all types of problems in 
quantum mechanics, simple from the physical point of view, and its 
correct application results in the exact energy eigenvalues for {\em 
all} solvable potentials.

In spite of long history no any strict rules concerning the 
application of the WKB method to multi-dimensional problems in 
quantum mechanics have been formulated. Meanwhile, this topic is 
closely related to the problem of exactness of the WKB method. The 
exactness of the method has proven in the literature for many 
potentials with the help of specially developed techniques, or 
improvements, or modifications of the quasiclassical method on the 
real axis and in the complex plane.  In this work we have fulfilled 
the quasiclassical analysis of the three-dimensional Schr\"odinger's 
equation. The original equation has been reduced to the form of the 
classic Hamilton-Jacobi equation without first derivatives. 
Separation of the equation has been performed with the help of the 
correspondence principle between classic and quantum-mechanical 
quantities. As a result of the separation, we have obtained the 
system of reduced second-order differential equations. Each of these 
equations has the correct quantum-mechanical form, $\hat 
p^2_q\psi(q)= p^2(q)\psi(q)$, and solved by the WKB method. We have 
stressed that the squared generalized moments, $p^2(q)$, obtained 
after separation should coincide with the corresponding classic 
moments. This means that the problem under consideration should 
correspond to a concrete classic problem.

We have shown that the Langer replacement $l(l+1)\rightarrow (l+\frac 
12)^2$ needed to reproduce the exact energy spectrum for the 
spherically symmetric potentials by the WKB method requires the 
modification of the squared angular momentum in the quasiclassical 
region. The squared angular momentum eigenvalues, $\vec M^2=(l+\frac 
12)^2\hbar^2$, have obtained in our approach from solution of the 
angular wave equation in the framework of the same quasiclassical 
method. As a result, the centrifugal term has the form $(l+\frac 
12)^2 \hbar^2/ r^2$ for {\em any} spherically symmetric potential 
$V(r)$.

The quasiclassical solution contains a more detail information in 
comparison with known exact solution. One of consequences of the WKB 
solution of the Schr\"odinger's equation is the existence of a 
nontrivial angular eigenfunction of the type of a standing half-wave 
for the angular quantum number $l=0$. This solution has treated as 
one which describes the quantum fluctuations of the angular momentum 
with the eigenvalue $M_0=\hbar/2$. We have shown that the quantum 
fluctuations of the angular momentum contribute to the energy of the 
ground state, $E_0$.

To demonstrate efficiency of the quasiclassical method, we have 
solved the three-dimensional Schr\"odinger's equation for some 
central-field potentials. The quasiclassical method successfully 
reproduces the exact energy spectrum not only for solvable 
spherically symmetric potentials but, also, for ``insoluble'' 
potentials with more than two turning points. The quasiclassical 
eigenfunctions for the discrete spectrum have written in elementary 
functions in the form of a standing wave.

The remarkable features of the quasiclassical method incline us to 
treat the leading-order WKB approximation as a special (asymptotic) 
exact method to solve the Schr\"odinger equation. In the 
quasiclassical approach we use the same technique for all types of 
problems. The same simple rules formulated for two-turning problems 
work for many turning point problems, as well. In this sense, the 
quasiclassical method is a more general in comparison with 
traditional one with the use of techniques of the special functions.

{\it Acknowledgements}.  The author thanks Prof. Uday P. Sukhatme for 
kind invitation to visit the University of Illinois at Chicago where 
a part of the work has been done and, also, for useful discussions 
and valuable comments. I should also like to thank Prof. A.A. Bogush 
for support and constant interest to this work.

This work was supported in part by the Belarusian Fund for 
Fundamental Researches.

\end{document}